\documentclass{elsarticle}
\usepackage{times}
\usepackage{amsfonts}
\usepackage{mathptmx}
\usepackage{bm}
\usepackage{graphicx,color}
\newcommand{\beq}{\begin{equation}}
\newcommand{\eeq}{\end{equation}}
\newcommand{\beqa}{\begin{eqnarray}}
\newcommand{\eeqa}{\end{eqnarray}}
\newcommand{\nn}{\nonumber \\}

\def \e {\mathrm{e}}

\def \el {\mathrm{el}}

\def \imb {\mathrm{imb}}

\def \la {\langle}
\def \ra {\rangle}
\def \s {\sigma}
\def \t {\tau}

\def \P {{\mathcal P}}

\def \Z {{\mathbb Z}}
\def \ch {\mathrm{ch}}

\def \z {\zeta}
\def \L {\underline{\Lambda}}    \def \D {\Delta}

\def \PF {\mathrm{PF}}
\def \Im {\mathrm{Im} \, }

\def \mod {\ \mathrm{mod} \ }
\def \H {{\mathcal H}}
\def \uu {{\widehat{u(1)}}}

\begin{document}
\begin{frontmatter}
\title{Thermopower and thermoelectric power factor of $\Z_k$ parafermion quantum dots}
\author{Lachezar S. Georgiev}
\ead{lgeorg@inrne.bas.bg}
\address{Institute for Nuclear Research and Nuclear Energy, Bulgarian Academy of Sciences,
	72 Tsarigradsko Chaussee, 1784 Sofia, Bulgaria}

\begin{abstract}
Using the conformal field theory approach to the thermoelectric characteristics of fractional quantum Hall states, previously 
developed in Nucl. Phys. B 894 (2015) 284, we show that the thermoelectric power factor of Coulomb-blockaded islands, 
realized by point contacts in Fabry--P\'erot interferometers in the $\Z_k$ parafermion Hall states, could give reliable signatures 
for distinguishing the topological orders of different quantum Hall states having identical electric properties. For example, while the 
conductance peak patterns in the Coulomb blockade regime for such states
 are practically indistinguishable for $v_n \ll v_c$ even at finite temperature, where $v_n$ and $v_c$ are the Fermi velocities 
of the neutral and charged modes respectively, the power factors $\P_T$ of the corresponding states are much more sensitive 
to the neutral modes. In particular, the smaller $r=v_n/v_c$ the bigger the asymmetries in the power factor which combined with 
the thermal broadening of the conductance peaks due to the neutral modes' multiplicities could give us the ultimate tool 
to figure out which of the competing quantum Hall universality classes are indeed realized in the experiments. 
We give a complete description of the power factor profiles in the $\Z_3$ and $\Z_4$ parafermion states with arbitrary 
number of quasiparticles localized in the bulk which could be useful for comparison with the experiments.
\end{abstract}
\end{frontmatter}
\section{Introduction}
The chiral edge excitations characterizing the topological order of the fractional quantum Hall (FQH) universality classes 
have been successfully described by 
conformal field theories (CFT) \cite{wen,fro-ker,ctz,read-CFT}. Some of these CFT models possess excitations with fractional electric 
charge and non-Abelian exchange statistics \cite{mr,rr} which makes them very promising in the field of topologically protected quantum 
information processing \cite{sarma-freedman-nayak,sarma-RMP}. Therefore it is very important to find experimentally observable 
signatures distinguishing between different universality classes of strongly interacting two-dimensional electron systems in strong 
perpendicular magnetic fields.
Usually for a given filling factor $\nu_H$ there are more than one candidate state, respectively
more than one CFT describing the corresponding universality class. Some of them are Abelian and share the same electric properties, 
such as the quantized Hall conductance, spectra of the electric charge of the quasiparticle excitations and conductance of a 
Coulomb blockaded quantum dot at low temperatures \cite{nayak-doppel-CB}. 
These states could still be distinguished by their thermoelectric properties, 
such as the thermal conductance, thermopower, figure of merit and thermoelectric power factor \cite{NPB2015}.
The role of thermopower for detecting non-Abelian quantum Hall states in the Corbino geometry
has been pointed out in Refs.~\cite{TP-PRB2009,TP-PRB2012}.

In this paper we will show how to apply the conformal field theory approach, developed in Ref.~\cite{NPB2015} within the linear 
response regime for Coulomb blockaded fractional quantum Hall states, to the calculation of experimentally important 
thermoelectric characteristics the $\Z_k$ parafermion quantum Hall states, i.e.,  how to use  the CFT partition function for these states
 in order to calculate the thermopower for a Coulomb blockaded quantum dot (QD), at non-zero temperature  for the experimental 
setup of Refs.~\cite{thermal,viola-stern,NPB2015}. 
We shall also assume that the reader is familiar with the diagonal coset construction of the $\Z_k$ parafermion 
quantum Hall states developed in Ref.~\cite{NPB2001}, as well as with the technical details of the Aharonov--Bohm flux incorporation 
into the disk CFT partition functions for FQH droplets, which was derived in Ref.~\cite{NPB-PF_k}. In this sense, the present work is 
a continuation of Ref.~\cite{NPB2015}, adding two more examples, the $\Z_3$  and $\Z_4$ parafermion Hall states, to the general 
CFT approach to the thermoelectric characteristics of Coulomb-blockaded FQH states, developed in Ref.~\cite{NPB2015}.

Measuring the power factor \cite{NPB2015} computed from the thermopower could experimentally
allow us to estimate the ratio $r=v_n/v_c$ of the Fermi velocities of the neutral and charged edge modes.
Notice that this ratio  might depend on the details of the experimental setup and might differ from sample to sample. 
Initially we consider the case $v_n=v_c$ when the low-energy effective field theory Hilbert space of the edge states of the QD
has a conformal symmetry, which allows us to write explicitly the Grand canonical partition function.
Bearing in mind that interactions could renormalize the Fermi velocities of the charged and neutral modes we can
compute various thermoelectric properties from this partition function, taking into account that if $v_n < v_c$  this changes 
only one parameter $\t \to \t'=r \t$
in the neutral sector of the above partition function \cite{NPB2015}. Comparing 
them with the corresponding quantities measured experimentally could eventually allow us to
distinguish between the different  candidate  quantum Hall states for the filling factor 
\beq \label{nu}
\nu_H=\frac{k}{kM+2}, \quad k=1, 2, \ldots, \quad M=\mathrm{odd}
\eeq
 such as the $\Z_k$ parafermion 
FQH states of Read--Rezayi \cite{rr} and their maximally symmetric \cite{fro-stu-thi} Abelian parents with $su(k) \oplus su(k)$ symmetry 
 introduced in Ref.~\cite{NPB2001} where the former is realized as a diagonal coset projection of the latter. In this paper we consider
 the most interesting case for which $M=1$.

The first member of the Read--Rezayi hierarchy of FQH states, corresponding to $k=2$, is the well known Moore--Read (Pfaffian)
state \cite{mr,NPB2001} whose thermoelectric properties have been investigated  in previous 
work \cite{viola-stern,NPB2015}. 
Here we will extend the analysis of Ref.~\cite{NPB2015}  to the other Read--Rezayi states in order to compute the 
thermopower of a Coulomb blockaded island in these FQH states in terms of the Grand-canonical averages of the 
edge states' Hamiltonian and particle number operators. To this end we will use the structure of the Grand-canonical 
partition functions for general FQH states on a disk properly modified in presence of Aharonov--Bohm (AB) flux \cite{NPB-PF_k}.
 We calculate numerically and plot the thermopower, the electric and thermal conductances 
and the power factors for these states with odd or even number of bulk quasiparticles. 

Following Ref.~\cite{NPB2015} we will express the thermopower $S$ (or, the Seebeck coefficient) and the corresponding power
factor $\P_T$
\beq \label{S}
S =-\frac{\la \varepsilon \ra}{eT}, \quad \P_T= S^2 G,
\eeq
where $G$ is the electric conductance of the island,
in terms of the average energy of the tunneling electrons $\varepsilon $ which can be computed from the Grand
canonical partition function of the edge states as the difference of the total QD energies with $N+1$ and $N$
electrons on the edge \cite{NPB2015}
\beq\label{eps}
\la \varepsilon \ra^{\phi}_{\beta,\mu_N} =
\frac{\la H_{\mathrm{CFT}}(\phi)\ra_{\beta,\mu_{N+1}} - \la H_{\mathrm{CFT}}(\phi)\ra_{\beta,\mu_N}}
{\la N_\el(\phi)\ra_{\beta,\mu_{N+1}} - \la N_\el(\phi)\ra_{\beta,\mu_N}}.
\eeq
Here $\beta=(k_BT)^{-1}$ is the inverse temperature,  $\la N_\el(\phi)\ra_{\beta,\mu_{N}}$ is the electron number thermal 
average on the edge,  with $N$ electrons at zero gate voltage (characterized by the chemical potential $\mu_N$),
as a function of the external side-gate voltage $V_g$ or AB flux $\phi$.
The variation of the side-gate voltage $V_g$ induces (continuously varying) ``external charge''
$e N_g=C_g V_g$ on the edge \cite{kouwenhoven,staring-CB},  which is equivalent to the AB flux-induced variation of the particle number 
$N_\phi=\nu_H \phi$, so that we can use instead of the gate voltage $V_g$ the equivalent  AB flux $\phi$ determined from
 \beq \label{V-phi}
C_gV_g/e\equiv \nu_H\phi \quad  \mathrm{with} \quad \phi=(e/h)\left( BA-B_0A_0\right),
\eeq
where $A_0$ is the area of the CB island and $B_0$ is the magnetic field at $V_g=0$. 

The electron number average on the QD's edge can be computed from the Grand potential $\Omega=-k_B T \ln Z(T,\mu)$
\beq \label{N}
\la N_\el (\phi)\ra_{\beta,\mu} = -\frac{\partial \Omega_{\phi}(T,\mu)}{\partial \mu} +\nu_H \phi ,
\eeq
and the Grand potential $\Omega_\phi$ in presence of non-zero side-gate voltage is expressed by
$\Omega_\phi=-k_B T \ln Z_\phi(T,\mu)$ with $Z_\phi(T,\mu)$ being the Grand partition function in presence of side-gate 
voltage \cite{NPB-PF_k}
\beq \label{Z2}
Z_{\phi}(\beta,\mu)= \mathop{\mathrm{tr}}_{\mbox{\quad}\H} \e^{-\beta (H_{\mathrm{CFT}}(\phi) -\mu N_{\imb})}.
\eeq
Here $\H$ is the Hilbert space at zero gate voltage, corresponding to $\phi=0$, the thermodynamic parameters $\beta$ and $\mu$ are
 independent of $\phi$, and all flux dependence is moved to the \textit{twisted
operator of energy} $H_{\mathrm{CFT}}(\phi)$ and the charge imbalance \cite{staring-CB} $N_{\imb}$ 
(cf. Eqs.~(32) and (33) in \cite{NPB-PF_k})
\beq\label{H-N}
H_{\mathrm{CFT}}(\phi)=H_{\mathrm{CFT}} -\Delta\varepsilon \phi N_\el +\frac{\nu_H}{2} \Delta\varepsilon\phi^2, \quad 
N_{\imb}=N_\el -\nu_H \phi ,
\eeq
where $N_\el$ and $H_{\mathrm{CFT}}$ are the electron number operator and the CFT Hamiltonian, respectively, at zero gate voltage.
The ultimate effect of the AB flux on the partition function $Z(\beta,\mu)$  is shifting 
$\mu \to \mu+\phi\Delta \varepsilon$, i.e.,  $Z_\phi(\beta,\mu) = Z(\beta,\mu+\phi\Delta \varepsilon)$, where 
$\Delta \varepsilon =\hbar 2 \pi v_F /L$ is the non-interacting energy spacing of electrons with Fermi velocity $v_F$ on
a QD of circumference $L$.

The conductance $G$ of the Coulomb blockaded island in presence of AB flux   (or gate voltage),
which is needed for the computation of the power factor $\P_T$, 
can be computed at finite temperature (in the linear response regime) according to Eq.~(10) in \cite{thermal}  
with $\mu=0$ 
\beq \label{G}
G (\phi)=\frac{e^2}{h}
\left( \nu_H +\frac{1}{2\pi^2} \left(\frac{T}{T_0} \right)\frac{\partial^2 }{\partial \phi^2}  \ln Z_{\phi}(T,0)\right) 
\eeq
and is zero in large intervals called CB valleys while showing sharp peaks at certain positions of the gate voltage
as illustrated in Fig.~\ref{fig:TP-G-Z3-00-T10-r1} for the $\Z_3$ parafermion FQH state. The temperature scale $T_0$
in Eq.~(\ref{G}) is defined through $k_B T_0 = \Delta  \varepsilon /2\pi$.

Notice that Eq.~(\ref{G}) gives only the conductance of the CB island. The total conductance of the Fabry--P\'erot interferometer,
or the single-electron transistor (SET) realized with the help of a side gate as in Fig. 1 in Ref.~\cite{NPB2015}, is \cite{thermal}
\[
G_{\mathrm{SET}} =\frac{h}{e^2}\frac{G_LG_R}{G_L+G_R}G(\phi),
\]
where $G_L$ and $G_R$ are the tunneling conductances of the left and right quantum point-contacts, which are assumed 
to be energy independent in the linear response regime but might depend on the temperature \cite{thermal}.

Finally, in order to compute the CB island's thermopower by Eq.~(\ref{eps}), we calculate the average quantum dot energies 
with $N$ electrons on the edge at temperature $T$ and 
chemical potential $\mu$ in presence of AB flux from the standard Grand 
canonical ensemble relation \cite{kubo,NPB2015}
\beq \label{H_CFT}
\la H_{\mathrm{CFT}}(\phi)\ra_{\beta,\mu} =\Omega_{\phi}(\beta,\mu)- T \frac{\partial \Omega_{\phi}(\beta,\mu)}{\partial T}  
 - \mu\frac{\partial  \Omega_{\phi}(\beta,\mu)}{\partial \mu}.
\eeq
As a final detail of the recipe of Ref.~\cite{NPB2015} for the computation of the thermopower through Eq.~(\ref{eps}) we emphasize 
that the chemical potentials 
$\mu_N$ and $\mu_{N+1}$ of the QD with $N$ and $N+1$ electrons respectively should be chosen as \cite{NPB2015}
\[
\mu_{N}=-\frac{\Delta \varepsilon}{2}, \quad \mu_{N+1}=\frac{\Delta \varepsilon}{2}.
\]
\section{Partition functions for  the $\Z_k$ parafermion FQH states}
\label{sec:part-func}
One interesting connection has been found in Ref.~\cite{NPB2001} between the $\Z_k$ Read--Rezayi FQH states, which 
have been originally introduced \cite{rr} as trial  many-body wave functions of $k+1$ -body $\delta$ function interaction 
Hamiltonians,
and the maximally-symmetric chiral quantum Hall (MSCQH) lattices in the classification of Ref.~\cite{fro-stu-thi}.
The MSCQH lattices describe Abelian CFTs which are rational extensions of the $\uu^{N}$
current algebra \cite{CFT-book} with a finite number  of $N$-dimensional vertex operators whose charges form an odd integer 
$N$-dimensional lattice with positive definite Gram matrix \cite{fro-stu-thi}. The simplest MSCQH lattice, denoted by the symbol
$(M+2 \; | \; {}^{\L_1}A_{k-1} \ {}^{\L_1}A_{k-1})$ in the notation of Ref.~\cite{fro-stu-thi}, which reproduces
the filling factor (\ref{nu}) has been shown \cite{NPB2001} to contain a $\uu$ current algebra corresponding to a Luttinger liquid
 with compactification radius $k(k+2)$ (for the case $M=1$) and  two copies of the $\widehat{su(k)}_1$  current algebra, corresponding
 to lattices whose metrics is equal to the Cartan matrix of the Lie algebra $A_{k-1}$,  altogether  combined in a non-trivial way
 by $\Z_k$ pairing rules  \cite{NPB2001}.

Although the  MSCQH lattice corresponding to $(3 \; | \; {}^{\L_1}A_{k-1} \ {}^{\L_1}A_{k-1})$ is not decomposable, due to the 
non-trivial weights $\L_1$ which coincide with the first $su(k)$ fundamental weight,   it does contain
 a decomposable sublattice corresponding to the symbol  $(k(k+2) \; | \; {}^{0}A_{k-1} \ {}^{0}A_{k-1})$ and the two lattices 
 can be related by \textit{gluing relations}  \cite{NPB2001}. This allows to express the Hilbert space corresponding to the original
 MSCQH lattice as a direct sum of tensor products of the Hilbert spaces corresponding to the decomposable sublattice 
\cite{NPB2001} which are irreducible representations of $\uu \oplus \widehat{su(k)}_1 \oplus \widehat{su(k)}_1$. Then the $\Z_k$
parafermions are obtained by a \textit{diagonal} coset projection \cite{GKO,CFT-book,NPB2001} in the neutral sector corresponding to
the removal of the diagonal subalgebra $\widehat{su(k)}_2$
\beq \label{PF_k}
\widehat{su(k)}_1 \oplus \widehat{su(k)}_1 \to \frac{\widehat{su(k)}_1 \oplus \widehat{su(k)}_1}{\widehat{su(k)}_2} \equiv \PF_k.
\eeq
The central charge of the diagonal coset is equal to the difference of the central charges of the CFTs 
for the numerator and denominator of the coset
 \cite{NPB2001,CFT-book}
\beq \label{c_PF}
c^{\PF}=2(k-1) - 2\frac{k^2-1}{k+2} =\frac{2(k-1)}{k+2}.
\eeq
Finally, the chiral partition functions for the edge states, which are defined as traces of the Boltzmann operator 
$\mathrm{e}^{-\beta (H-\mu N)}$
over the above mentioned Hilbert spaces, can be written as \cite{NPB2001}
\beq \label{full-ch}
\chi_{l,\rho} (\t,\z) = \sum_{s=0}^{k-1} K_{l+s(k+2)}(\t,k\z;k(k+2)) \ch(\L_{l-\rho+s}+\L_{\rho+s})(\t),
\eeq
where  $K_{l}(\t,\z; m)$ is the $\uu$ partition function  for the charged part which is completely determined by the 
filling factor $\nu_H$ and 
coincides with that for a chiral Luttinger liquid with a compactification radius \cite{cz} $R_c=1/m$, in the 
notation of  \cite{CFT-book,NPB-PF_k} 
 \beq \label{K}
K_{l}(\t,\z; m) = \frac{\mathrm{CZ}}{\eta(\t)} \sum_{n=-\infty}^{\infty} q^{\frac{m}{2}\left(n+\frac{l}{m}\right)^2} 
\e^{2\pi i \z \left(n+\frac{l}{m}\right)}.
\eeq
Here  $q=\e^{2\pi i \t }=\e^{-\beta\Delta\varepsilon}$, where $\beta=(k_B T)^{-1}$ is the inverse temperature,   
\[
\eta(\t)=q^{1/24}\prod_{n=1}^\infty (1-q^n)
\]
 is the Dedekind function \cite{CFT-book} and 
 $\mathrm{CZ}(\t,\z)=\exp(-\pi\nu_H(\Im \z)^2/\Im\t)$ is the Cappelli--Zemba factor  needed to preserve the 
invariance of $K_{l}(\t,\z; m)$ with respect to the Laughlin spectral flow \cite{cz}.

The modular parameter $\z$ used in the definition of the rational CFT partition functions is related to the chemical potential $\mu$
by \cite{NPB-PF_k,NPB2015}
\beq \label{zeta}
\z= \frac{\mu}{\Delta\varepsilon} \t, \quad \mathrm{where} \quad  \tau=i\pi\frac{T_0}{T}, \quad T_0=\frac{\hbar v_c}{\pi k_B L}
\eeq
and transforms after introducing AB flux $\phi$ as \cite{NPB2001,NPB2015}
\beq \label{zeta-phi}
\z \to \z+\phi \t \quad \Longleftrightarrow \quad \mu \to \mu +\phi \Delta \varepsilon,
\eeq
which leads only to the following transformation of the $\uu$ partition functions \cite{NPB2001,NPB2015}
\beq\label{K-phi-mu}
K_l(\t,k\z;k(k+2)) \to K_l(\t,k(\z+\phi\t);k(k+2)) \equiv  K_{l+k(\phi+\mu/\Delta\varepsilon)} (\t,0;k(k+2)).
\eeq
The right-hand side of Eq.~(\ref{K-phi-mu}) is convenient for numerical calculations because the $K$ function for $\z=0$ is real.

The range of admissible labels $l $ is $\mod k+2$ while the range of $\rho$  is $\mod k$ subjected to the condition
$l-\rho \leq \rho \mod k$ \cite{NPB2001}.

The neutral partition functions  $\ch(\L_{\mu}+\L_{\rho})(\t)$ of the diagonal coset $\PF_k$ are labeled 
in principle by an admissible weight for the current algebra $\widehat{su(k)}_2$, which can be written as a sum of two 
fundamental $su(k)$ weights, i.e., $\L_{\mu}+\L_{\rho}$ with $0\leq \mu \leq \rho \leq k-1$.
Following Refs.~\cite{lep-primc,schill1,schill2} these characters for the diagonal coset CFT can be written as
\beq\label{5.18}
\ch_{\s,Q}(\t;\PF_k)= q^{ \D^\PF(\s) - \frac{c^\PF}{24}  }
\sum\limits_{
\mathop{m_1,m_2,\ldots, m_{k-1}=0}\limits_{\sum\limits_{i=1}^{k-1}\, i
\, m_i \equiv Q \mod k } }^\infty
\frac{q^{\underline{m}. C^{-1}.
\left(\underline{m} - \L_\s \right)} }{(q)_{m_1} \cdots (q)_{m_{k-1}} },
\qquad
(q)_n=\prod\limits_{j=1}^n (1-q^j)
\eeq
where  $\underline{m} = (m_1,\ldots, m_{k-1})$ is a $k-1$ component vector with 
non-negative integer components  in the basis of $su(k)$ fundamental weights $\{ \L_1,\ldots \L_{k-1}  \}$,
$\D^\PF(\L_\s)=\s(k-\s)/(2k(k+2))$ is the CFT dimension of the primary field characterized by the coset triple  \cite{NPB2001} 
of weights
$(\L_\s,0;\L_0+\L_\s)$, for $\L_\s \in \{0,\L_1,\ldots, \L_{k-1} \}$, the central charge $c^\PF$ is given by Eq. (\ref{c_PF})
and $C^{-1}$ is the  inverse $su(k)$ Cartan matrix.
All the sectors could be obtained uniquely \cite{NPB2001} if
$0\leq \s \leq Q \leq k-1$. The restriction
$\sum\limits_{i=1}^{k-1}\  i \, m_i \equiv Q \mod k $ implements a projector
on the irreducible representation of $\PF_k$ with a fixed $\Z_k$-charge which we can now specify.
It is worth mentioning that the characters $\ch_{\s,Q}\equiv \ch(\L_\mu+\L_\rho)$ derived in Eq.~(\ref{5.18}) are the true 
characters \cite{NPB2001} of the diagonal coset $\PF_k$  labeled  in the standard way by the level-2 weights
$\L_\mu+\L_\rho$, where $0\leq \mu \leq \rho \leq k-1$. Then the parameters $(\s,Q)$ are related to $(\mu, \rho)$ by \cite{NPB2001}
\beq \label{mu-rho}
\s=\rho-\mu, \quad Q=\rho  \quad \Longleftrightarrow \quad
\mu=Q-\s, \quad  \rho= Q .
\eeq
Now that we know the relation between the $0\leq \s \leq Q \leq k-1$ indices and $0\leq \mu \leq \rho \leq k-1$
we can determine \cite{NPB2001}  the $\Z_k$ charge of the characters $\ch_{\s}^Q$
\beq \label{Z_k-charge}
P = \mu+\rho \mod k= 2Q-\s \mod k.
\eeq
We have to mention that the characters formulae  (\ref{5.18}) are
of the type of the {\it Universal Chiral Partition Function} (UCPF)
introduced by Berkovich and McCoy \cite{ucpf}
and used by them and Schoutens \cite{schout-spinons-PLB,schout-spinons-NPB,schout1,schout2} for the analysis of the
exclusion statistics in the FQH effect.

The first member of the $\Z_k$ parafermion hierarchy, corresponding to $k=2$, is the Moore--Read (Pfaffian) FQH state.
Its thermoelectric properties have been investigated in Ref.~\cite{viola-stern,NPB2015}. In particular, the thermopower 
and the power factor for the cases with even and odd number of bulk quasiparticles have been computed numerically and plotted in
\cite{NPB2015}. In the next Section we consider the next member of the hierarchy, corresponding to $k=3$.
\section{Example 1:  the $\Z_3$ parafermion FQH state}
As an interesting illustration of the approach of Ref.~\cite{NPB2015} we consider a CB island in which the  FQH state is in the $\Z_3$ 
Read--Rezayi (parafermion) state \cite{rr,NPB2001}, characterized by   $n_H=3$, $d_H=5$, i.e. $\nu_H=3/5$.

The $\Z_3$ parafermion  (Read--Rezayi) state is the most promising candidate to describe the experimentally 
observed \cite{pan,xia,pan-xia-08}
incompressible state in the second Landau level, corresponding to filling factors $\nu_H= 13/5 = 2+3/5$ (not very well developed 
in the experiments\footnote{the diagonal resistance $R_L$ of the incompressible quantum Hall states should vanish for zero temperature 
at the plateaus of the off-diagonal resistance $R_H$, however, at non-zero temperature there are instead  local minima of $R_L$, 
which tend to zero when $T\to 0$. The minima of $R_L$ for the $\nu_H=12/5$ plateau are more developed in the experiments 
\cite{pan,xia,pan-xia-08} than the minima of  $R_L$ at $\nu_H=13/5$, so that it is believed that there is a FQH state at $\nu_H=12/5$,
 while the existence of a FQH plateau at $\nu_H=13/5$ is still questionable.}) 
and its particle--hole conjugate at  $\nu_H= 12/5 = 3-3/5$. The most appealing characteristics of this state is that it possesses non-Abelian
quasiparticle excitations, which are topologically equivalent  to the Fibonacci anyons \cite{fibonaci,sarma-RMP}. These anyons  can be used to 
construct multielectron wave functions which belong to degenerate manifolds whose dimension increases exponentially with the number 
of anyons and can be used for \textit{universal} topologically protected quantum information processing \cite{fibonaci,sarma-RMP}.
While the $\nu_H=12/5$ FQH state is less stable than the more popular $\nu_H=5/2$ FQH state, which is believed to be described by
the Majorana fermion of the Ising model, the latter is known to be not universal \cite{clifford}, in the sense that not all elementary 
quantum operations can be 
implemented by braiding Ising anyons \cite{TQC-NPB,sarma-RMP} and are therefore not topologically protected. 
On the contrary, all elementary quantum gates in the Fibonacci quantum computer can be implemented
by braiding of Fibonacci anyons \cite{fibonaci} and all they are fully protected from noise and decoherence by 
the topology of the quantum computer \cite{sarma-RMP}. 
Therefore, it is really challenging to find whether the observed FQH state at $\nu_H=12/5$ is indeed
the $\Z_3$ parafermion state. Since there are more states, corresponding to the same filling factor and with the same electric characteristics 
like the $\Z_3$ parafermion state, belonging to different universality classes, including Abelian ones 
(i.e., such that do not possess non-Abelian quasiparticle excitations) it is important to find measurable quantities which can 
distinguish between the different candidates. Because those candidates have different neutral components, one way to 
distinguish between them is to measure physical characteristics, such as the thermopower and the  thermoelectric power factor 
considered in this paper, 
which are sensitive to minor changes in the neutral degrees of freedom of the strongly correlated two-dimensional electron system.

Numerical evidence suggests that the Jain FQH state corresponding to $\nu_H=2/5$ is unstable in the second Landau level 
\cite{jain-CF2-0,jain-CF2},
so there are not many candidates to describe the observed FQH state at $\nu_H=2+2/5$. Yet, there is at least one more candidate 
state to describe the filling factors $\nu_H=2+3/5$ and $3-3/5=2+2/5$ which is the MSCQH lattice state, described in 
Section~\ref{sec:part-func}, which is the Abelian parent \cite{NPB2001} of the  parafermion coset (\ref{PF_k}) characterized by the 
$\widehat{su(3)}_1\oplus \widehat{su(3)}_1$ symmetry.

The properties of the edge states in the $\Z_3$ parafermion island depend on the type and number of quasiparticles localized in the bulk.
There are 10 topologically inequivalent quasiparticle excitations \cite{NPB2001}, corresponding to 10 different disk partition functions 
of the type (\ref{full-ch}). However, there are only two different  patterns of CB conductance peaks, 
corresponding to no quasiparticles in the bulk and one quasiparticle localized in the bulk. 
All other cases are equivalent to one of these two plus a translation on the horizontal axis
corresponding to some additional quanta of the AB flux.
\subsection{No quasiparticles in the bulk}
The disk partition function for the $\Z_3$ FQH state without quasiparticles in the bulk,  which corresponds to $l=0$, $\L=0$ 
in Eq.~(\ref{full-ch}) takes the form \cite{NPB2001}
\beqa
Z_{0,0}(\t,\z)&=&K_0(\t,3\z;15) \mathrm{ch}_{0,0}(r\t) +K_5(\t,3\z;15) \mathrm{ch}_{0,1}(r\t) \nn
 &+&K_{-5}(\t,3\z;15) \mathrm{ch}_{0,2}(r\t) 
\eeqa
where $r=v_n/v_c$,  the $K$ functions are defined in Eq.~(\ref{K}) and the  neutral partition functions are defined by \cite{NPB2001}
\[
\ch_{0,l}(\t) = q^{-\frac{1}{30}} \sum_{n_1, \ n_2\geq 0 }^{(l)}
\frac{q^{\frac{2}{3}\left(n_1^2+n_1n_2+n_2^2 \right)}}{(q)_{n_1}(q)_{n_2}}, 
(q)_n=\prod_{j=1}^n (1-q^j) .
\]
Notice  that the sum $\sum^{(l)}$ is restricted by the condition
 ${n_1+2n_2=l \mod 3}$.
 
\begin{figure}[htb]
\centering
\includegraphics[bb=50 15 560 390,clip,width=10cm]{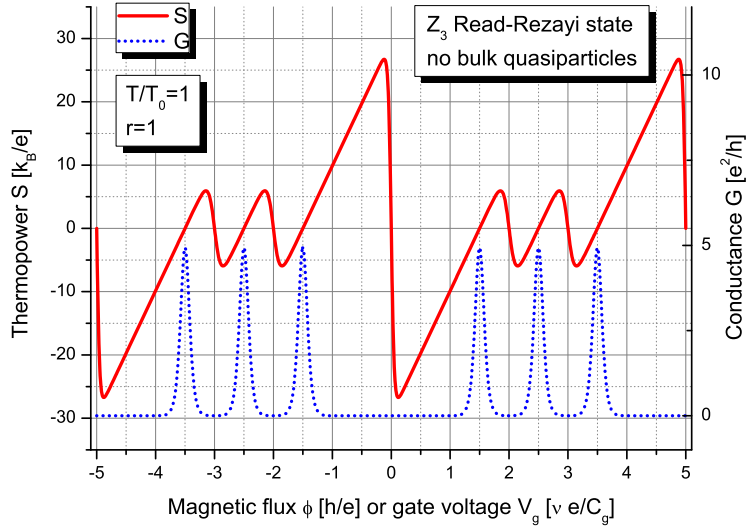}
\caption{Thermopower and electric conductance for a CB island in the $\Z_3$-parafermion (Read-Rezayi) state 
with $\nu_H=2+3/5$, 
without quasiparticles in the bulk,  with $r=1$, at temperature $T/T_0=1$. \label{fig:TP-G-Z3-00-T10-r1}}
\end{figure}

After introducing AB flux as in Eq.~(\ref{zeta-phi}) and using the property (\ref{K-phi-mu}) of the $K$ functions (\ref{K})
we can write the partition function as
\beqa\label{Z00}
Z_{0,0}^{\phi}(\t,\z)&=&K_{3(\phi+\mu/\Delta\varepsilon)}(\t,0;15) \mathrm{ch}_{0,0}(r\t) +
K_{5+3(\phi+\mu/\Delta\varepsilon)}(\t,0;15) \mathrm{ch}_{0,1}(r\t) \nn
 &+&K_{-5+3(\phi+\mu/\Delta\varepsilon)}(\t,0;15) \mathrm{ch}_{0,2}(r\t) 
\eeqa
and we calculate numerically the thermopower and the conductance of the CB island from Eqs.~(\ref{eps}),  
(\ref{N}), (\ref{H_CFT}) and (\ref{G}) at temperature $T/ T_0=1$ and $r=1$,  see Fig.~\ref{fig:TP-G-Z3-00-T10-r1}.
 Although the plot of the thermopower for $r=1$ might be unrealistic, as experiments and numerical calculations suggest that 
$v_n < v_c$, it is instructive for the analysis of the characteristics of the thermopower in general, which are similar to those 
for metallic islands: 
the thermopower grows linearly with the gate voltage $V_g$ and the edges are smoothened at finite temperature; it 
 is always 0 at the maximum of a conductance peak expressing the fact that the tunneling energy is zero at the conductance peaks; 
there are sharp jumps of thermopower (discontinuous at $T=0$)  in the middle of the CB valleys between neighboring 
conductance peaks, expressing particle-hole symmetry \cite{matveev-LNP}. 
The plot of the thermopower for the $\Z_3$ parafermion FQH state is similar to that of the Moore--Read (Pfaffian) state~\cite{NPB2015},
and to that of the superconducting SET~\cite{TP-SC} as well, 
only the number of small-amplitude oscillations of $S$ has been increased form one to two, corresponding to the number of short-period
conductance peaks. This allows us to anticipate the general pattern of the oscillations of the thermopower in the $\Z_k$ parafermion states
without quasiparticles localized in the bulk: it naturally confirms the results obtained in  Eq.~(4.18) of Ref.~\cite{cappelli-viola-zemba} 
and Eq.~(27) of Ref.~\cite{stern-CB-RR-PRB} for $r=1$
that the conductance peaks for these states are bunched in groups  of $k$ equidistant peaks 
(with distance between peaks equal to $\Delta=1$)
and the groups are separated by a larger distance $\Delta+2r =3$. For general $r<1$ the conductance peaks will be separated by
\beq \label{spacing-0}
(\Delta+2r,\underbrace{\Delta, \ldots, \Delta}_{k-1}) \quad \mathrm{for} \quad l=0 \ \mathrm{or} \ k
\eeq
so that there will be $k-1$ small-amplitude thermopower oscillations, corresponding to the conductance peaks separated by $\Delta$ and one 
large-amplitude oscillation corresponding to the larger spacing $\Delta+2r$.  Moreover, because the sum of all spacings between the conductance 
peak positions should be equal to the period $(k+2)$, we find a condition
\beq\label{spacing-1}
k\Delta +2r = k+2 \quad \Longrightarrow \quad \Delta=\frac{k+2 -2r}{k} \quad \Longleftrightarrow \quad r=\frac{k+2-k\Delta}{2}
\eeq
(for $0 \mod k$ quasiparticles in the bulk) from which we can express $\Delta$ in terms of $r$, or vice versa,
which is equivalent to the results of Refs.~\cite{stern-CB-RR-PRB,cappelli-viola-zemba}.
In all cases we expect that the maxima of the conductance 
peaks would correspond to the zero of thermopower with positive slope, while the (discontinuous) jumps, or thermopower 
zeros with negative slope -- to the centers of the CB valleys.

\begin{figure}[htb]
\centering
\includegraphics[bb=30 15 570 430,clip,width=12cm]{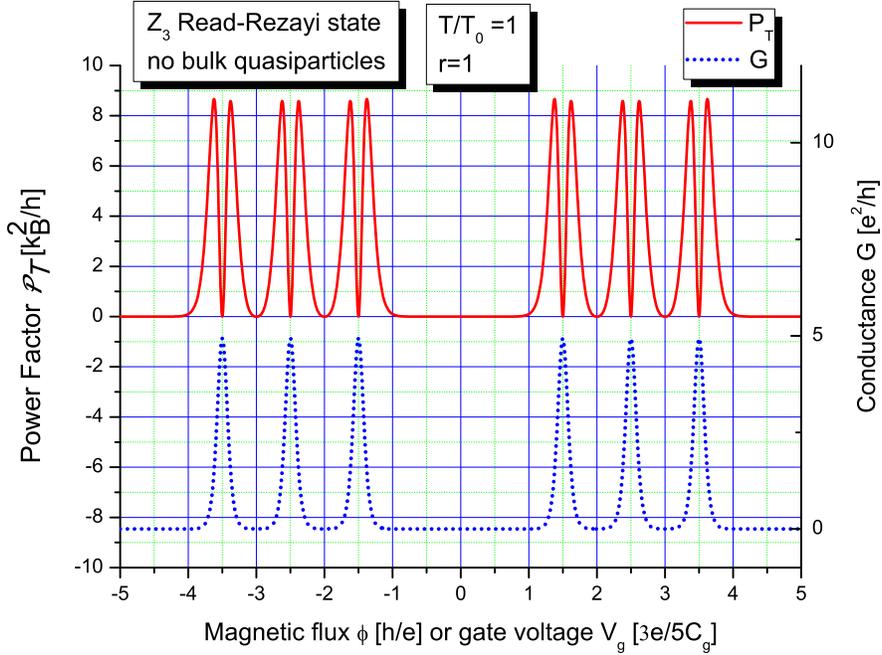}
\caption{Power factor (left Y-scale, straight red line) and electric conductance (right Y-scale, dotted blue line) for a CB island in 
the $\Z_3$-parafermion (Read-Rezayi) state with $\nu_H=2+3/5$, without quasiparticles in the bulk,  with $r=1$, 
at temperature $T/T_0=1$. \label{fig:PF-G-Z3-00-T10-r1}}
\end{figure}
In Fig.~\ref{fig:PF-G-Z3-00-T10-r1} we plot the power factor $\P_T$ of the $\Z_3$ parafermion state without quasiparticles in the 
bulk for $v_n=v_c$ at temperature $T=T_0$ together with the conductance $G$. As is obvious from Eqs.~(\ref{spacing-0}) and 
(\ref{spacing-1}) for $r=1$ the smaller period for $k=3$ is $\Delta=1$, while the larger one is $\Delta+2r =3$.

It is worth emphasizing that, as obvious from Fig.~\ref{fig:PF-G-Z3-00-T10-r1}, the power factor $\P_T$ possesses sharp dips 
corresponding precisely to the positions of the conductance peaks because then the thermopower $S$ vanishes, while the 
conductance $G$ is non-zero and the $\P_T$ depends quadratically on $S$, according to Eq.~(\ref{S}).
This fact makes the power factor $\P_T$ a perfect tool for measuring the positions of the conductance peaks and then using 
Eqs.~(\ref{spacing-0}) and (\ref{spacing-1}) for calculating the ratio $r=v_n/v_c$ of the velocities of the neutral and charged edge 
modes. As shown in Ref.~\cite{NPB2015} as well as in this paper, the profiles of the power factors when gate voltage is varied are 
highly sensitive to the value of $r$ and could be very asymmetric for different values of $r$ and in different FQH states. 
It is also very promising that the power factor $\P_T$ for a SET could be directly measurable  by the methods of 
Ref.~\cite{gurman-2-3} which opens
a wide experimental opportunity for distinguishing FQH states at low but finite temperature for realistic parameters of the samples.

\begin{figure}[htb]
\centering
\includegraphics[bb=20 20 560 460,clip,width=11cm]{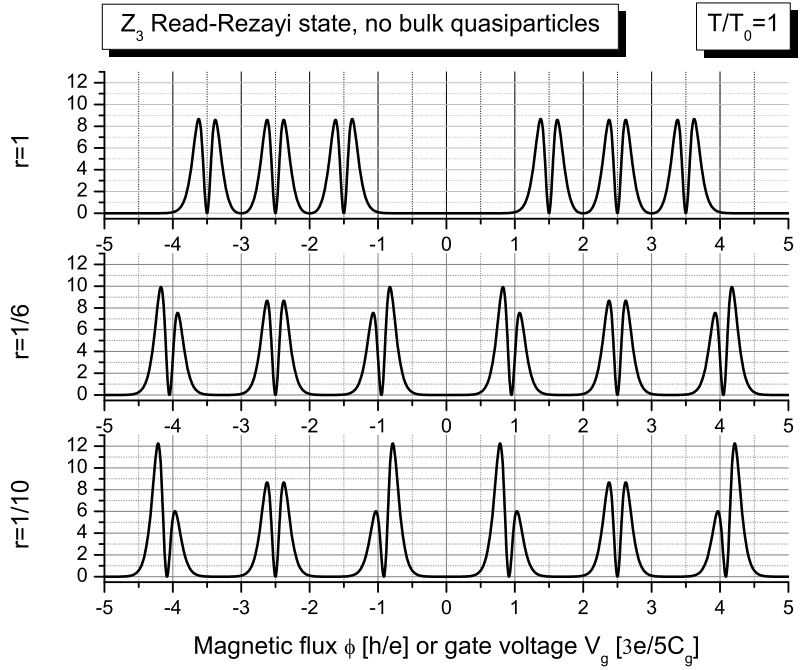}
\caption{Power factors' $\P_T$ profiles in units [$k_B^2/h$] for a CB island in the $\Z_3$-parafermion (Read-Rezayi) 
state with $\nu_H=2+3/5$, 
without quasiparticles in the bulk,  with $r=1$, $r=1/6$ and $r=1/10$ at temperature $T/T_0=1$. \label{fig:PF-Z3-00-T10-r1-r1-6-r1-10}}
\end{figure}

In Fig.~\ref{fig:PF-Z3-00-T10-r1-r1-6-r1-10} we compare the profiles of the power factor $\P_T$ for the $\Z_3$ parafermion FQH state 
without quasiparticles in the bulk at temperature $T=T_0$ for three different values of $r$: $r=1$, $r=1/6$ and  $r=1/10$.
Interestingly enough, the thermoelectric power factor $\P_T$ displays large asymmetries when $r$ is decreased, 
unlike the electric conductance patterns 
which become indistinguishable when $r$ decreases. It is also interesting that the power-factor's dips at some of the conductance peaks
(e.g. those at $\phi = \pm 2.5$) for the case without bulk quasiparticles, remain unchanged and symmetric for all three values 
of $r$ while the others become very asymmetric.

The difference between the thermopower of the $\Z_3$ parafermions and that of the metallic islands, due to the role of the neutral 
degrees of freedom in the CB island with non-Fermi liquid behavior,  
can be seen in the ratio between the maxima of the thermoelectric power factor at the neighboring conductance peaks which  
depends on the non-universal ratio $r=v_n/v_c$ and this observable dependence is rather sensitive to the FQH universality class 
and to the number of quasiparticles in the bulk of the CB islands. This result is similar to those for the thermoelectric 
power factor profiles for the $\nu_H=5/2$ quantum Hall states in CB islands \cite{NPB2015}. It was shown there that 
the ratio  between the maxima of the thermoelectric power factor at the neighboring conductance peaks is sensitive to the 
neutral multiplicities in the partition functions and the value of the non-universal parameter $r$ and their temperature
behavior could be used to distinguish between the different candidates for the $\nu_H=5/2$ plateau \cite{NPB2015}.
\subsection{One quasiparticle localized in the bulk}
Once again we notice that  in order to obtain the complete list of characters of
the $\Z_k$-parafermions it is sufficient to take only the fundamental
weights  $\L_\s$ for $\s=0,1,\ldots, k-1$ and to project onto subspaces
having definite $\Z_k$-charge determined by $P$.
For example, in the case
$k=3$ we have $\L_\s \in \left\{ (0,0),(1,0),(0,1)\right\} $.
Then the other sectors are
reproduced by the action of the polynomials in the negative modes of
$(2,0)$ and $(0,2)$ -- the $\Z_3$ parafermionic currents of dimension $2/3$.
In particular, the sector with dimension $2/5$ is generated by the action of
$(2,0)$ (or $(0,2)$) on the sectors defined by $(1,0)$  (or $(0,1)$). 

The disk partition function for a CB island containing $\Z_3$ parafermion FQH state with one fundamental quasiparticle 
localized in the bulk \cite{NPB2001} 
of the island, in presence of AB flux $\phi$ or gate voltage $V_g$,  can be described by the following disk partition function, 
corresponding to $l=1$ and $\rho=1$ in Eq.~(\ref{full-ch}), according to Eqs.~(\ref{zeta-phi}) and (\ref{K-phi-mu})
\beqa \label{Z11}
Z_{1,1}^{\phi}(\tau,\mu) &=&K_{1+3(\phi+\mu)}(\t,0;15)\ch_{1,1}(r\t)+ K_{6+3(\phi+\mu)}(\t,0;15)\ch_{1,2}(r\t)+ \nn
& &K_{-4+3(\phi+\mu)}(\t,0;15)\ch_{2,2}(r\t)
\eeqa
where the $K$ functions are defined in Eq.~(\ref{K}), $r=v_n/v_c$ and the neutral partition functions are \cite{NPB2001}
\[
\ch_{1,1}(\t)= \ch(\L_0+\L_1;\t)=  q^{\frac{1}{30}} \sum_{\begin{array}{c} n_1, n_2\geq 0 \\ n_1+2n_2 \equiv 1 \mod 3\end{array}}
\frac{q^{\frac{2}{3}\left(n_1^2+n_1 n_2+n_2^2 \right)-\frac{2n_1+n_2}{3}}}{(q)_{n_1}(q)_{n_2}}, 
\]
\[
\ch_{2,2}(\t)=\ch(\L_0+\L_2;\t)= q^{\frac{1}{30}} \sum_{\begin{array}{c} \small { n_1}, n_2\geq 0 \\ n_1+2n_2 \equiv 2 \mod 3\end{array}}
\frac{q^{\frac{2}{3}\left(n_1^2+n_1 n_2+n_2^2 \right)-\frac{n_1+2 n_2}{3}}}{(q)_{n_1}(q)_{n_2}}, 
\]
\[
\ch_{1,2}(\t)=\ch(\L_1+\L_2;\t)=q^{\frac{1}{30}} \sum_{\begin{array}{c} n_1, n_2\geq 0 \\ n_1+2n_2 \equiv 2 \mod 3\end{array}}
\frac{q^{\frac{2}{3}\left(n_1^2+n_1 n_2+n_2^2 \right)-\frac{2n_1+n_2}{3}}}{(q)_{n_1}(q)_{n_2}}.
\]
\begin{figure}[htb]
\centering
\includegraphics[bb=20 10 560 440,clip,width=12cm]{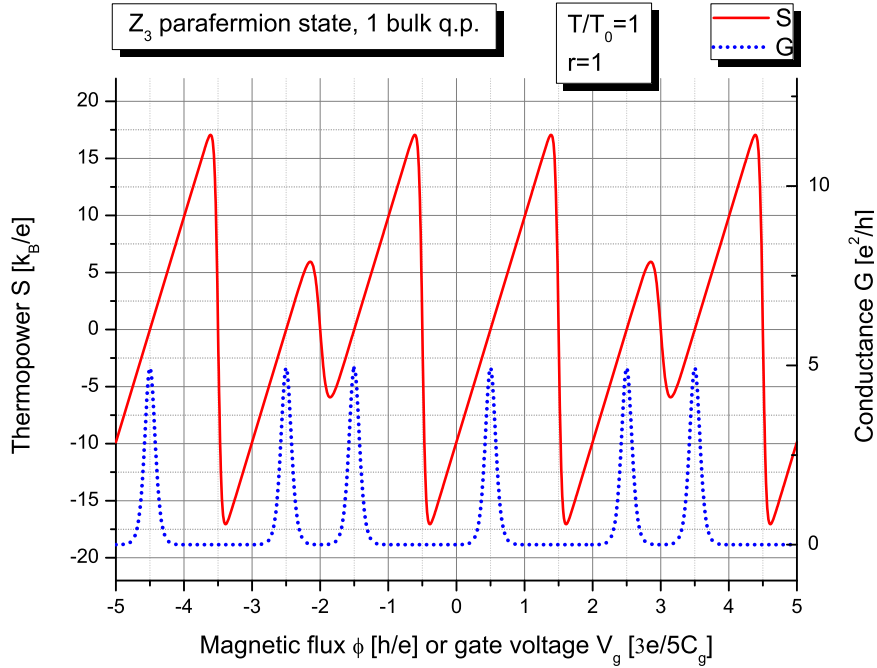}
\caption{Thermopower and electric conductance for a CB island in the $\Z_3$-parafermion (Read-Rezayi) state with $\nu_H=3/5$, 
with $1$ quasiparticle in the bulk,  with $r=1$, at temperature $T/T_0=1$. \label{fig:TP-G-Z3-11-T10-r1}}
\end{figure}
Having specified all partition functions entering Eq.~(\ref{Z11}) we plot in Fig.~\ref{fig:TP-G-Z3-11-T10-r1}  the oscillations of 
the thermopower $S$ and the conductance of the $\Z_3$ 
parafermion  island with one quasiparticle localized in the bulk, calculated from  Eqs.~(\ref{eps}),  
(\ref{N}), (\ref{H_CFT}) and (\ref{G}) at temperature $T/ T_0=1$ and $r=1$.
Again, like in Fig.~\ref{fig:TP-G-Z3-00-T10-r1}, there are small-amplitude oscillations of the thermopower $S$, 
corresponding to the short-period spacing between the conductance peaks and larger-amplitude oscillations corresponding 
to the longer-period spacing between the peaks, however, now the larger-amplitudes are smaller than in 
Fig.~\ref{fig:TP-G-Z3-00-T10-r1}. Still the zeros of the thermopower correspond to the maxima of the conductance peaks 
and the discontinuities correspond to the centers of the CB valleys \cite{matveev-LNP}.
Yet, the pattern of the thermopower oscillations is different-- while in Fig.~\ref{fig:TP-G-Z3-00-T10-r1} the pattern of oscillations is 
``small-small-big-small-small-big'', here the patter is ``big-big-small-big-big-small''.

\begin{figure}[htb]
\centering
\includegraphics[bb=10 15 570 460,clip,width=11cm]{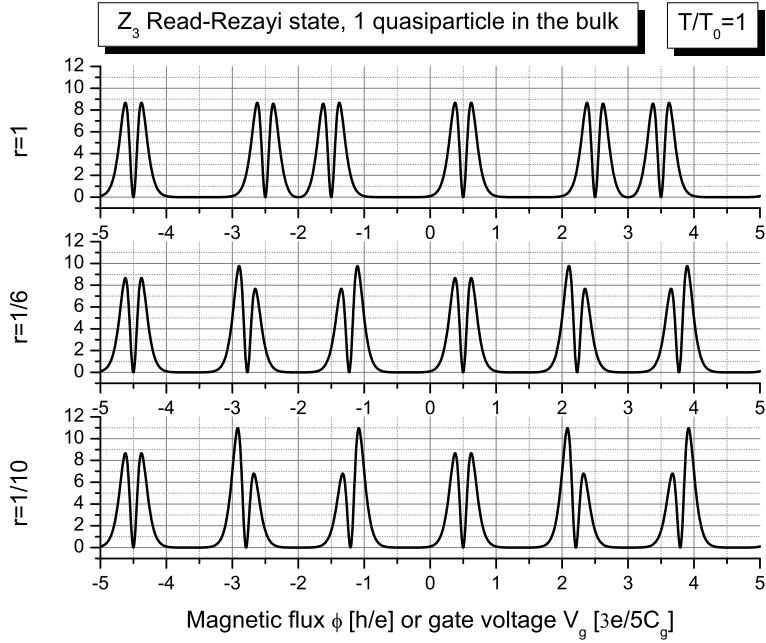}
\caption{Power factors in units $[k_B^2/h]$ for a CB island in 
the $\Z_3$-parafermion (Read-Rezayi) state with $\nu_H=3/5$, with 1 quasiparticle in the bulk, corresponding to $l=1$ 
and $\rho=1$ in Eq.~(\ref{full-ch}) for $k=3$,   with $r=1$, $r=1/6$ and $r=1/10$ 
at temperature $T/T_0=1$. \label{fig:PF-Z3-11-T10-r1-r1-6-r1-10}}
\end{figure}

Looking at Fig.~\ref{fig:TP-G-Z3-11-T10-r1}  it is not difficult to anticipate the basic profile of the power factor $\P_T$
for the $\Z_3$ parafermion state with $r=1$: each conductance peak will correspond to a pair of sharp peaks of $\P_T$
which are symmetric and the sharp dip between them points out the exact position of the peak 
(see, the uppermost plot in Fig.~\ref{fig:PF-Z3-11-T10-r1-r1-6-r1-10}). However, when the ratio $r$ is decreased the power factor
$\P_T$ displays rather noticeable asymmetries. Therefore, in Fig.~\ref{fig:PF-Z3-11-T10-r1-r1-6-r1-10} we plot the power factors $\P_T$ 
for the $\Z_3$ parafermion state with 1 quasiparticle localized 
in the bulk for different values of $r$:  $r=1$, $r=1/6$ and $r=1/10$.
Again, like in the case without quasiparticles in the bulk,  there are two CB peaks positions, those  at $\phi=-4.5$ and $\phi=0.5$, 
which do not change in shape and size when we change $r$. 
The exact positions of and the spacing between the conductance peaks can be obtained from the sharp dips of the power factor
for any $r$.
Both the positions and spacings depend weakly on the ratio $r$ and are difficult to be extracted from the conductance data alone.
As shown in Eq.~(4.18) in Ref.~\cite{cappelli-viola-zemba} and in Eqs.~(25) and (26) in Ref.~\cite{stern-CB-RR-PRB} for $l=1$
and $k=3$ the conductance peaks form two groups of equidistant peaks (with spacing $\Delta$ between the peaks in the group): 
one group containing $l=1$ peaks and one group containing $k-l=2$ (equidistant) peaks  and the groups are separated by the 
larger spacing $\Delta+r$, i.e., 
\beq \label{spacing-2}
(\Delta+r,\Delta,\Delta+r) \quad \mathrm{for} \quad l=1 \ \mathrm{and} \ k=3 .
\eeq
Again the sum of all spacings should give the period $k+2$ so we arrive at the same relation between $\Delta$ and $r$ like 
Eq.~(\ref{spacing-1}) for $k=3$.
\subsection{Two quasiparticles localized in the bulk}
Because of the non-Abelian statistics of the quasiparticles in the $\Z_3$ parafermion FQH state, which is expressed in the 
fusion rule \cite{NPB2001}
\beq\label{fusion-2qp}
(\L_0+\L_1) \times (\L_0+\L_1) \simeq (\L_0+\L_2) + (\L_1+\L_1) ,
\eeq
there are two distinct disk partition functions with two quasiparticles in presence of AB flux $\phi$ and non-zero chemical 
potential $\mu$: one corresponding to $l=2$  and $\rho=1$ in
Eq.~(\ref{full-ch}) for $k=3$
\beqa \label{Z21}
Z_{2,1}^{\phi}(\tau,\mu) &=&K_{2+3(\phi+\mu)}(\t,0;15)\ch_{2,2}(r\t)+ K_{7+3(\phi+\mu)}(\t,0;15)\ch_{1,1}(r\t)+ \nn
& &K_{-3+3(\phi+\mu)}(\t,0;15)\ch_{1,2}(r\t)
\eeqa
and the second corresponding to $l=2$  and $\rho=2$ in
Eq.~(\ref{full-ch}) for $k=3$
\beqa \label{Z22}
Z_{2,2}^{\phi}(\tau,\mu) &=&K_{1+3(\phi+\mu)}(\t,0;15)\ch_{0,1}(r\t)+ K_{6+3(\phi+\mu)}(\t,0;15)\ch_{0,2}(r\t)+ \nn
& &K_{-4+3(\phi+\mu)}(\t,0;15)\ch_{0,0}(r\t) .
\eeqa
The plot of the power factors obtained from the partition function $Z_{2,1}$
for three different values $r=1$, $r=1/6$ and $r=1/10$ at temperature $T/T_0=1$
is the same as Fig.~\ref{fig:PF-Z3-11-T10-r1-r1-6-r1-10} for a single quasiparticle localized in the bulk,
except that the graph is shifted with 1 flux  quantum to the left on the horizontal axis.

Similarly, the plot of the power factors obtained from the partition function $Z_{2,2}$
for three different values $r=1$, $r=1/6$ and $r=1/10$ at temperature $T/T_0=1$
is the same as Fig.~\ref{fig:PF-Z3-00-T10-r1-r1-6-r1-10} for the case without quasiparticles localized in the bulk,
except that the graph is shifted to the right with 1 quantum of flux on the horizontal axis.
\subsection{Three quasiparticles localized in the bulk}
When we have three non-Abelian quasiparticles localized in the bulk of the $\Z_3$ parafermion disk state,
 which is expressed by the fusion rule \cite{NPB2001}
\[
(\L_0+\L_1) \times (\L_0+\L_1) \times (\L_0+\L_1)\simeq (\L_0+\L_0) + (\L_1+\L_2) ,
\]
there are two distinct disk partition functions with three quasiparticles: one corresponding to $l=0$  and $\rho=0$,
i.e., no quasiparticles in the bulk which we have specified in Eq.~(\ref{Z00}), 
and another corresponding to $l=0$  and $\rho=2$, in Eq.~(\ref{full-ch}) for $k=3$
which can be written as
\beqa \label{Z02}
Z_{0,2}^{\phi}(\tau,\mu) &=&K_{3(\phi+\mu)}(\t,0;15)\ch_{1,2}(r\t)+ K_{5+3(\phi+\mu)}(\t,0;15)\ch_{2,2}(r\t)+ \nn
& &K_{-5+3(\phi+\mu)}(\t,0;15)\ch_{1,1}(r\t).
\eeqa
The plot of the power factors obtained from the partition function $Z_{0,2}$
for three different values $r=1$, $r=1/6$ and $r=1/10$ at temperature $T/T_0=1$
is the same as Fig.~(\ref{fig:PF-Z3-11-T10-r1-r1-6-r1-10}) for a single quasiparticle localized in the bulk,
except that the graph is shifted with 2 flux  quantum to the right on the horizontal axis.

To conclude this section we summarize that the thermopower $S$ and the corresponding power factor $\P_T$
for the $\Z_3$ parafermion island depend on the number of quasiparticles localized in the bulk, however,
all patterns are similar to those of the cases without bulk quasiparticles and one bulk quasiparticle plus 
some translation on the horizontal axis.
%
\section{Example 2:  the $\Z_4$ parafermion FQH state}
The next member of the $\Z_k$ parafermion hierarchy corresponds to $k=4$ and we recall \cite{NPB2001} 
that its neutral sector is described by the $\Z_4$ 
parafermions of Ref.~\cite{zf-para}. The $\Z_4$ parafermion FQH state is a candidate to describe the experimentally observed
Hall states at filling factors $\nu_H=2+2/3$ and its particle--hole conjugate at $\nu_H=3-2/3$ \cite{pan,xia,pan-xia-08}.
The central charge of the parafermion part of the CFT is $c^\PF=2(k-1)/(k+2)=1$. Numerical and experimental work suggests that this state
is competing with the Laughlin state and its particle--hole conjugate  in the second Landau level \cite{xia,pan-xia-08} 
and transitions between these states are possible under certain conditions. 
Another candidate state is the MSCQH lattice state \cite{NPB2001}, 
described in Section~\ref{sec:part-func}, which is the Abelian parent 
of the  parafermion coset (\ref{PF_k}) characterized by the $\widehat{su(4)}_1\oplus \widehat{su(4)}_1$ symmetry.
While the Laughlin states are known to be more stable than the others \cite{jain-CF2}, the investigation of the non-Abelian 
$\Z_4$ parafermion state is important for the completeness of the phase diagram and possible applications to the 
topological quantum computation \cite{sarma-RMP}.
\subsection{No quasiparticles localized in the bulk}
\label{sec:Z4_00}
The disk partition function for the $\Z_4$ parafermion FQH state without quasiparticles in the bulk corresponds to
$l=0$ and $\rho=0$ in Eq.~(\ref{full-ch}) and can be written as
\beqa \label{Z4-00}
Z_{0,0}^{\phi}(\t,\mu)=
K_{4(\phi+\mu)}(\t,0;24) \ch_{0,0}(r\t) + K_{6+4(\phi+\mu)}(\t,0;24)\ch_{0,1}(r\t) + \nn
K_{12+4(\phi+\mu)}(\t,0;24)\ch_{0,2}(r\t) + K_{-6+4(\phi+\mu)}(\t,0;24)\ch_{0,3}(r\t)
\eeqa
$r=v_n/v_c$, the $K$ functions are given in Eq.~(\ref{K}), we have used Eq.~(\ref{K-phi-mu}) to move the dependence on 
$\phi$ and $\mu$ into the indices of the $K$ functions and we denoted $\mu/\Delta\varepsilon=:\mu$. 
The neutral characters of the $\Z_4$ parafermion CFT  are given by
\beq \label{ch-neutral}
\ch_{\s,Q}(\t)=q^{\frac{\s(4-\s)}{48}-\frac{1}{24}}
\sum_{n_1\ge 0, n_2\ge 0, n_3 \ge 0} p(n_1,n_2,n_3;Q)
\frac{q^{\Delta(n_1,n_2,n_3;\s)}}{(q)_{n_1} (q)_{n_2} (q)_{n_3} }
\eeq
where $0\leq \s \leq Q\leq 3$, the $q$-polynomial $(q)_{n}$ is defined in Eq.~(\ref{5.18}) and we introduced an explicit projector
\[
p(n_1,n_2,n_3;Q)=\left\{ \begin{array}{l}  1 \quad \mathrm{iff} \quad n_1+2n_2+3n_3=Q \mod 4 \cr
0 \quad \mathrm{otherwise}
\end{array} \right.  .
\]
The dimension $\Delta(n_1,n_2,n_3;\s)$ in Eq.~(\ref{ch-neutral}) can be written as a matrix product as
\[
\Delta(n_1,n_2,n_3;\s)=
\frac{1}{4}(n_1,n_2,n_3) \left( \matrix{3 & 2 & 1 \cr 2 & 4 & 2 \cr 1 & 2 & 3} \right) 
\left[\left( \begin{array}{c} n_1 \cr n_2 \cr n_3  \end{array}\right) - \L_\s \right]  ,
\]
where the transpose of the vector-column $\L_\s$  is the one in the set \\
$\{ [0,0,0],[1,0,0],[0,1,0],[0,0,1]\}$ whose $k$-ality  is $[m_1,m_2,m_3]= m_1+2m_2+3m_3  = \s \mod 4$.
\begin{figure}[htb]
\centering
\includegraphics[bb=40 15 560 410,clip,width=11cm]{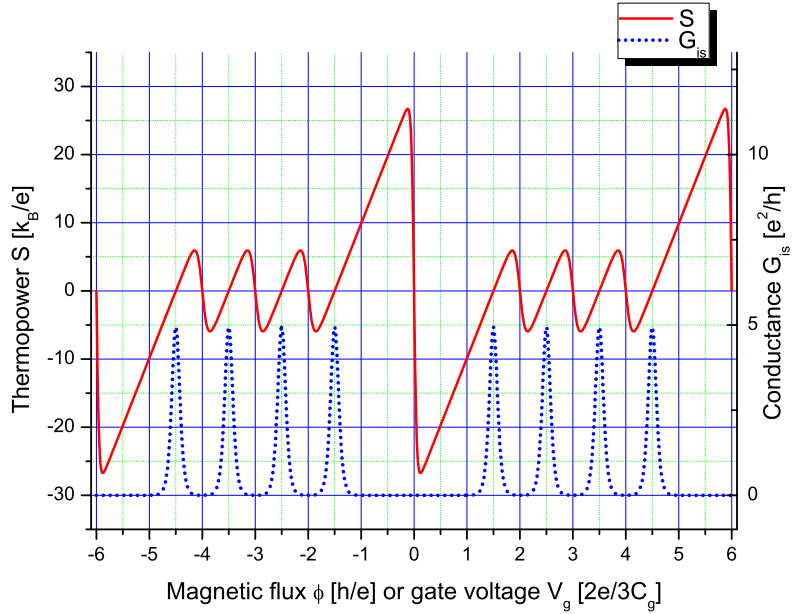}
\caption{Thermopower $S$ (left Y-scale, straight red line) and electric conductance $G$ (right Y-scale, blue dotted line) 
for a CB island in the $\Z_4$-parafermion (Read-Rezayi) state with $\nu_H=2/3$, 
without quasiparticles in the bulk,  with $r=1$, at temperature $T/T_0=1$. \label{fig:TP-G-Z4-00-T10-r1}}
\end{figure}
Choosing some upper limit for $n_i$ in Eq.~(\ref{ch-neutral}), e.g. $0 \leq n_1, n_2,n_3 \leq 10$, which is sufficient since 
$q\approx 2.7 \times 10^{-9}$ for $T=T_0$,  we calculate the neutral 
partition functions (\ref{ch-neutral}) numerically and plot in Fig.~\ref{fig:TP-G-Z4-00-T10-r1} the thermopower and the electric 
conductance of a $\Z_4$ parafermion island with $v_n=v_c$ at temperature $T=T_0$. 
Again, like in the $\Z_3$ parafermion state, the thermopower shows
3 small-amplitude oscillations, corresponding to the short-period $\Delta=1$ in the AB flux $\phi$ and one large-amplitude
oscillation, corresponding to the long-period $\Delta+2r=3$ in the flux. The zeros of the thermopower correspond to the 
maxima of the conductance peaks, expressing the fact that the tunneling is maximal when the difference in the energy 
of the quantum dots with $N$ and $N+1$ electrons vanishes. Also, there is a sharp jump (discontinuous at $T=0$) in
the middle of the CB valleys, i.e., for $\phi=0$, expressing the particle--hole symmetry at this position of the AB flux.
\begin{figure}[htb]
\centering
\includegraphics[bb=15 15 560 440,clip,width=11cm]{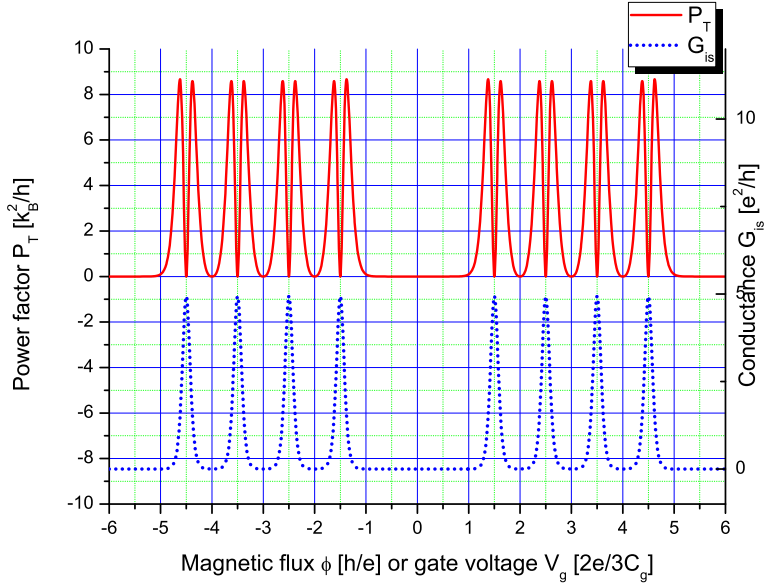}
\caption{Power factors $\P_T$ and electric conductance for a CB island in the $\Z_4$-parafermion (Read-Rezayi) 
state with $\nu_H=2/3$, without quasiparticles in the bulk, with $r=1$ at temperature $T/T_0=1$.
 \label{fig:PF-G-Z4-00-T10-r1}}
\end{figure}
 Next we plot in Fig.~\ref{fig:PF-G-Z4-00-T10-r1} the power factor $\P_T$ obtained from the thermopower $S$ according to
 Eq.~(\ref{S}), for the  $\Z_4$ parafermion island without bulk quasiparticles for $v_n=v_c$ at temperature $T=T_0$.
 Again, it is obvious that the positions of the conductance maxima precisely correspond to the sharp dips in the power factor 
$\P_T$ and this could be used to determine the conductance peak positions when $v_n< v_c$.

Finally we close the case of  the $\Z_4$-parafermion (Read-Rezayi)  state without quasiparticles in the bulk
by plotting the profiles of the power factors $\P_T$ at $T=T_0$ for three different values of $r=v_n/v_c$: 
$r=1$, $r=1/6$ and $r=1/10$. 
\begin{figure}[htb]
\centering
\includegraphics[bb=15 15 560 480,clip,width=11cm]{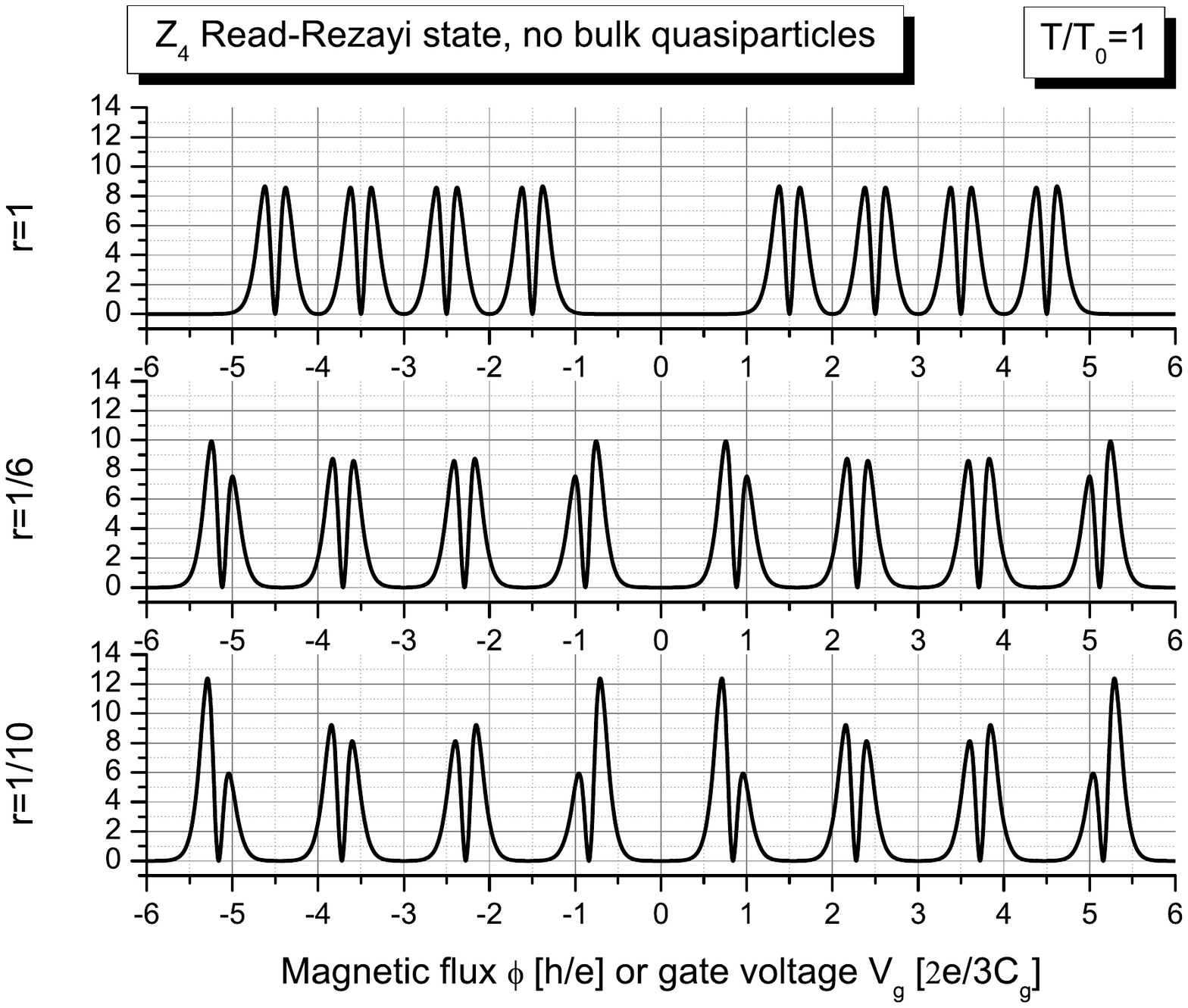}
\caption{Power factors in units [$k_B^2/h$] for a CB island in the $\Z_4$-parafermion (Read-Rezayi) state with $\nu_H=2/3$, 
without quasiparticles in the bulk,  with  $r=1$,  $r=1/6$ and $r=1/10$, at temperature 
$T/T_0=1$. \label{fig:PF-Z4-00-T10-r1-r1-6-r1-10}}
\end{figure}
 This plot shows that when $r$ is decreased the power factor $\P_T$ displays noticeable asymmetries in the heights of the 
$\P_T$ peaks surrounding the conductance peaks positions, which combined with the 
 relatively precise measurement of the ratio $r$, could give some experimental signature for distinguishing different candidates
to describe the experimental data. 
Following the approach of Ref.~\cite{NPB2015}  we can propose an experimental way to choose between the different candidates
for the $\nu_H=12/5$ Hall state: assume that the profile of the thermoelectric power factor for this FQH state has been measured 
experimentally, e.g., by the methods of Ref.~\cite{gurman-2-3} -- then from the dips of the measured power factor we can estimate the value 
of $r$ and recalculate the theoretical prediction for the power factor of the different candidate states for that $r$. After that we can 
compare the theoretical power factors with the experimental one and choose the closest one by the $\chi^2$ method.
\subsection{One quasiparticle localized in the bulk}
\label{sec:Z4_11}
The disk partition  function for the $\Z_4$ parafermion state with one quasiparticle localized in the bulk, which corresponds to $l=1$ and 
$\rho=1$ in Eq.~(\ref{full-ch}), scan be written in presence of AB flux $\phi$ and non-zero chemical potential $\mu$ as
\beqa \label{Z4_11}
Z_{1,1}^{\phi}(\t,\mu)=
K_{1+4(\phi+\mu)}(\t,0;24) \ch_{1,1}(r\t) + K_{7+4(\phi+\mu)}(\t,0;24)\ch_{1,2}(r\t) + \nn
K_{-11+4(\phi+\mu)}(\t,0;24)\ch_{2,2}(r\t) + K_{-5+4(\phi+\mu)}(\t,0;24)\ch_{3,3}(r\t),
\eeqa
where the  $K$ functions are defined in Eq.~(\ref{K}), $r=v_n/v_c$ and the neutral characters are defined in Eq.~(\ref{ch-neutral}).
\begin{figure}[htb]
\centering
\includegraphics[bb=40 15 560 440,clip,width=12cm]{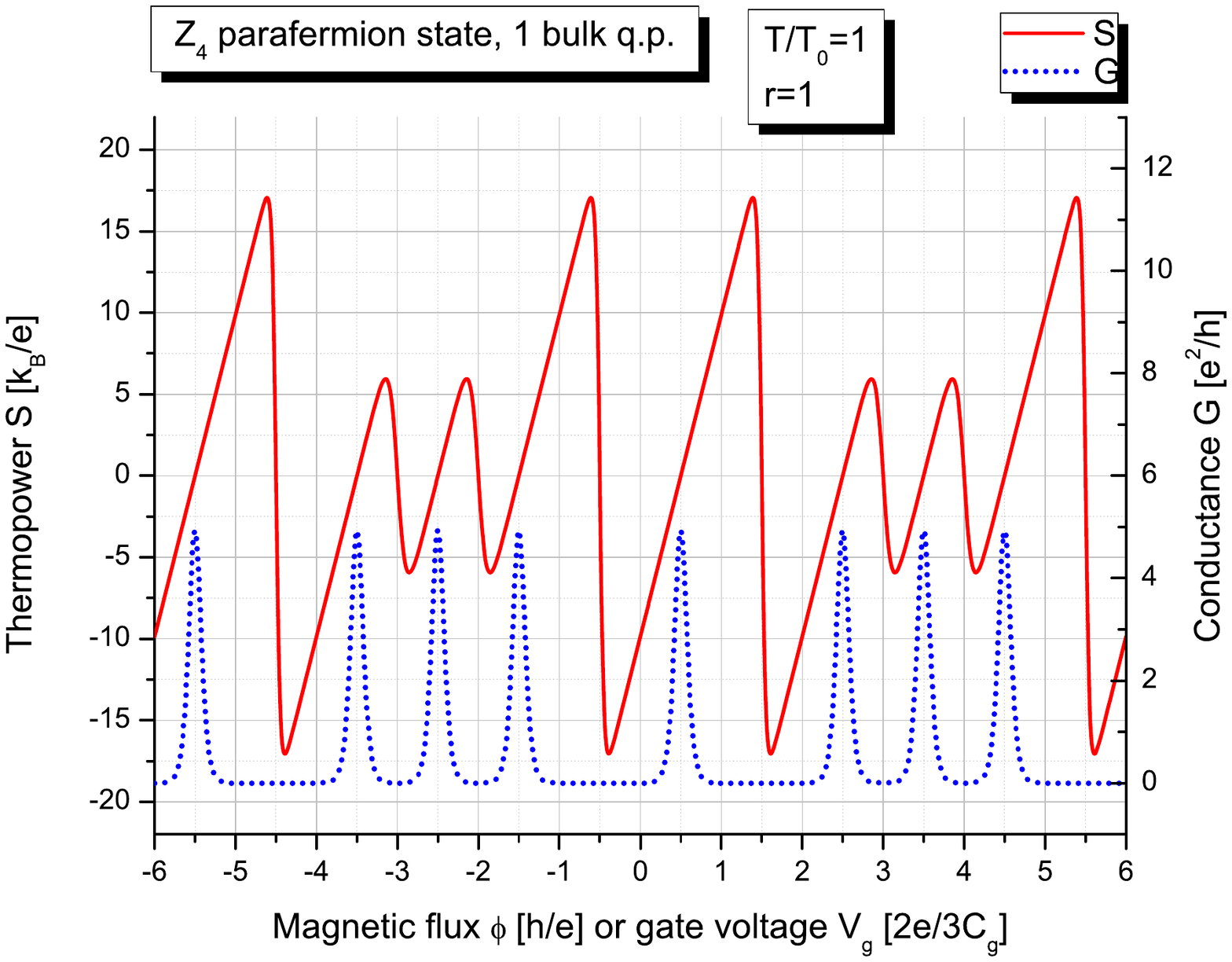}
\caption{Thermopower $S$ (left Y-scale, straight red line) and electric conductance $G$ (right Y-scale, blue dotted line) 
for a CB island in the $\Z_4$-parafermion (Read-Rezayi) state with $\nu_H=2/3$, 
with one quasiparticle in the bulk,  for $r=1$, at temperature $T/T_0=1$. \label{fig:TP-G-Z4-11-T10-r1}}
\end{figure}
Substituting the explicit partition function (\ref{Z4_11}), again with the restriction  $0 \leq n_1, n_2,n_3 \leq 10$ for the neutral 
characters and taking 200 terms in the $K$ functions (\ref{K}), we calculate numerically and plot in Fig.~\ref{fig:TP-G-Z4-11-T10-r1}
together the electric conductance $G$ and the thermopower $S$ for the $\Z_4$ parafermion state with one fundamental quasiparticle 
localized in the bulk for temperature $T=T_0$ and $v_n=v_c$.
 We see that again the zeros of the thermopower correspond to the maxima of the conductance peaks and 
the zero-temperature discontinuities of the thermopower correspond to the centers of the Coulomb blockade valleys.
Again the profiles of the thermopower and conductance peaks exactly reproduce the results of 
Eq.~(4.18) in Ref.~\cite{cappelli-viola-zemba} and Eqs.~(25) and (26) in Ref.~\cite{stern-CB-RR-PRB} showing that there are 
two groups of equidistant conductance peaks, one with $l=1$ peaks and the second with $k-l=3$ peaks, 
with in-group spacing $\Delta$, and the groups are separated by a larger spacing $\Delta+r$.
This corresponds to the following thermopower oscillation pattern:``small-small-big-big-small-small-big-big''
which is certainly distinguishable from the previous pattern for the case without bulk quasiparticles as long as there is an 
observable difference between small-amplitude and big-amplitude oscillations of $S$.

\begin{figure}[htb]
\centering
\includegraphics[bb=15 15 560 480,clip,width=11cm]{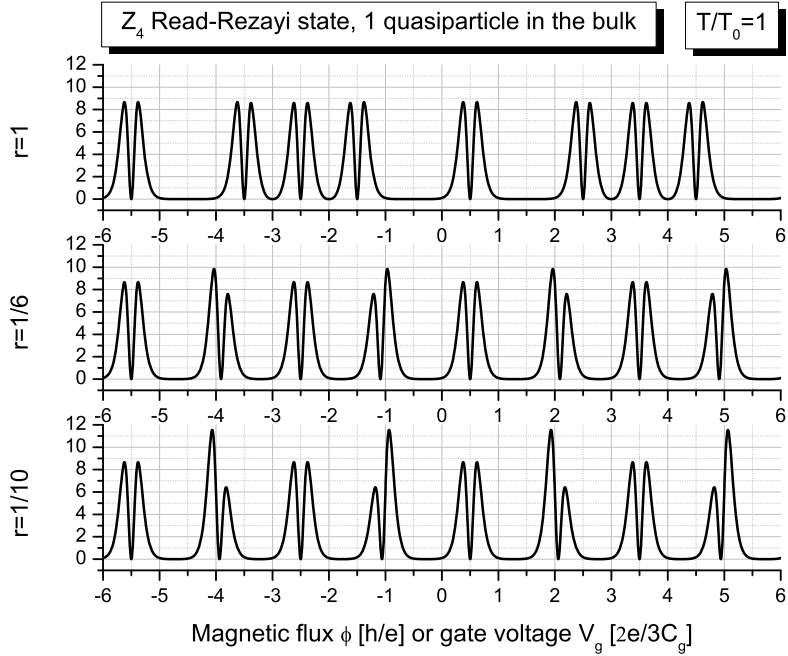}
\caption{Power factors in units [$k_B^2/h$] for a CB island in the $\Z_4$-parafermion (Read-Rezayi) state with $\nu_H=2/3$, 
with $1$ quasiparticle localized  in the bulk, corresponding to $l=1$ and $\rho=1$  in Eq.~(\ref{full-ch}), 
with  $r=1$,  $r=1/6$ and $r=1/10$, at temperature 
$T/T_0=1$. \label{fig:PF-Z4-11-T10-r1-r1-6-r1-10}}
\end{figure}

The thermoelectric power factor for $r=1$ is shown in the uppermost graph of 
Fig.~\ref{fig:PF-Z4-11-T10-r1-r1-6-r1-10} and is completely symmetric with the same spacing of the dips like the conductance 
peaks  in Fig.~\ref{fig:TP-G-Z4-11-T10-r1}. 
For comparison with the $\P_T$ profiles for other values of $r$ we also plot  in Fig.~\ref{fig:PF-Z4-11-T10-r1-r1-6-r1-10}  
the power factors $\P_T$ of the $\Z_4$-parafermion (Read-Rezayi)  state with one quasiparticle in the bulk for the other 
two values of $r=v_n/v_c$: $r=1/6$ and $r=1/10$ at $T=T_0$.
Again we see two conductance peaks (at $\phi=-2.5$ and  $\phi=3.5$) which give symmetric profile of the power factor  
and are unchanged for all values of $r$.
The plots  in Fig.~\ref{fig:PF-Z4-11-T10-r1-r1-6-r1-10} show that when $r$ is decreased the power factor $\P_T$ displays 
noticeable asymmetries in the heights of the $\P_T$ peaks surrounding the conductance peaks positions, which combined with the 
 relatively precise measurement of the ratio $r$, could give some experimental signature for distinguishing different candidates
to describe the experimental data, as discussed before. 
\subsection{Two or more quasiparticles localized in the bulk}
\label{sec:Z4_21}
\begin{figure}[htb]
\centering
\includegraphics[bb=40 15 580 440,clip,width=12cm]{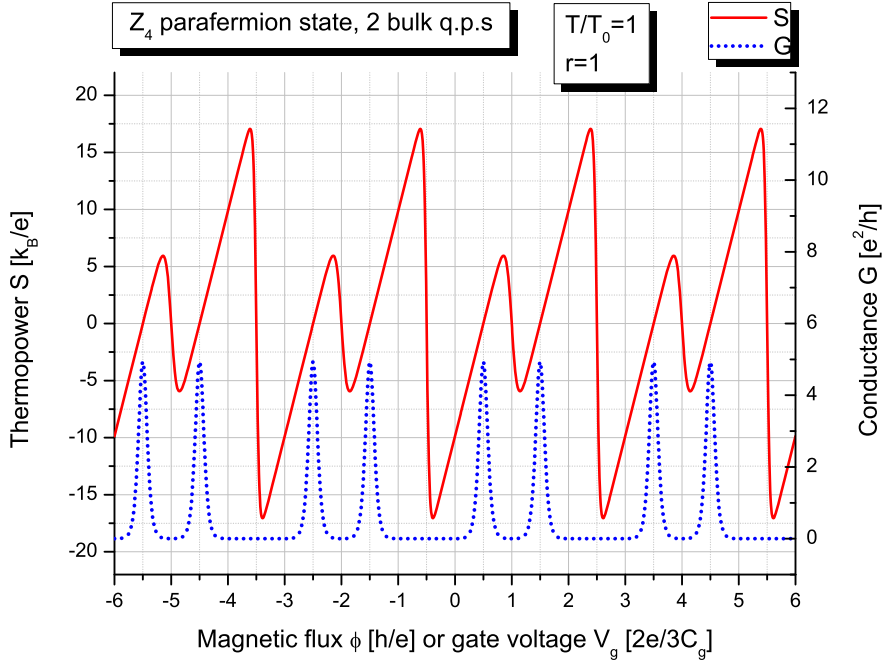}
\caption{Thermopower $S$ (left Y-scale, straight red line) and electric conductance $G$ (right Y-scale, blue dotted line) 
for a CB island in the $\Z_4$-parafermion (Read-Rezayi) state with $\nu_H=2/3$, 
with two quasiparticles in the bulk,  corresponding to $l=2$ and $\rho=1$ in Eq.~(\ref{full-ch}), for $r=1$ 
at temperature $T/T_0=1$. \label{fig:TP-G-Z4-21-T10-r1}}
\end{figure}

Again, like in the $k=3$ case, due to the non-Abelian fusion rules between two fundamental quasiparticles,  
 which take the same form as Eq.~(\ref{fusion-2qp}) we have two distinct partition functions corresponding to the two different fusion 
channels in Eq.~(\ref{fusion-2qp}).
One of them is the disk partition  function which corresponds to $l=2$ and 
$\rho=1$ in Eq.~(\ref{full-ch}), and can be written in presence of AB flux $\phi$ and non-zero chemical potential $\mu$ as
\beqa \label{Z4_21}
Z_{2,1}^{\phi}(\t,\mu)=
K_{2+4(\phi+\mu)}(\t,0;24) \ch_{2,2}(r\t) + K_{8+4(\phi+\mu)}(\t,0;24)\ch_{1,2}(r\t) + \nn
K_{-10+4(\phi+\mu)}(\t,0;24)\ch_{2,2}(r\t) + K_{-5+4(\phi+\mu)}(\t,0;24)\ch_{3,3}(r\t),
\eeqa
where the  $K$ functions are defined in Eq.~(\ref{K}), $r=v_n/v_c$ and the neutral characters are defined in Eq.~(\ref{ch-neutral}).
Using again the restriction  $0 \leq n_1, n_2,n_3 \leq 10$ for the neutral characters in Eq.~(\ref{Z4_21}) and taking again 
200 terms in the $K$  functions (\ref{K}), we compute numerically and plot in Fig.~\ref{fig:TP-G-Z4-21-T10-r1} the 
thermopower $S$ and the conductance for the case with two quasiparticles localized in the bulk.
Again we see that the zeros of the thermopower $S$ coincide with the maxima of the conductance peaks and the 
zero-temperature discontinuities  of $S$ correspond to the CB valleys \cite{matveev-LNP}.
Again the plot of the thermopower and conductance peaks exactly reproduce the results of 
Eq.~(4.18) in Ref.~\cite{cappelli-viola-zemba} and Eqs.~(25) and (26) in Ref.~\cite{stern-CB-RR-PRB}.
We see that there are  two groups of equidistant conductance peaks (with spacing $\Delta$ inside each group), 
one with $l=2$ peaks and the second with  $k-l=2$ peaks, separated by a larger spacing $\Delta+r$ between the groups.
 Notice that the periodicity of the conductance peaks and thermopower oscillations is halved $(k+2)/2=3$ because 
$k=4$ is even, as mentioned  in Ref.~\cite{cappelli-viola-zemba}.

The pattern of the thermopower oscillations is now ``small-big-small-big'' and is obviously different from those for the cases
 without bulk quasiparticles (Fig.~\ref{fig:TP-G-Z4-00-T10-r1})  and with one bulk 
quasiparticle  (Fig.~\ref{fig:TP-G-Z4-11-T10-r1}). 
 
\begin{figure}[htb]
\centering
\includegraphics[bb=15 15 560 480,clip,width=11cm]{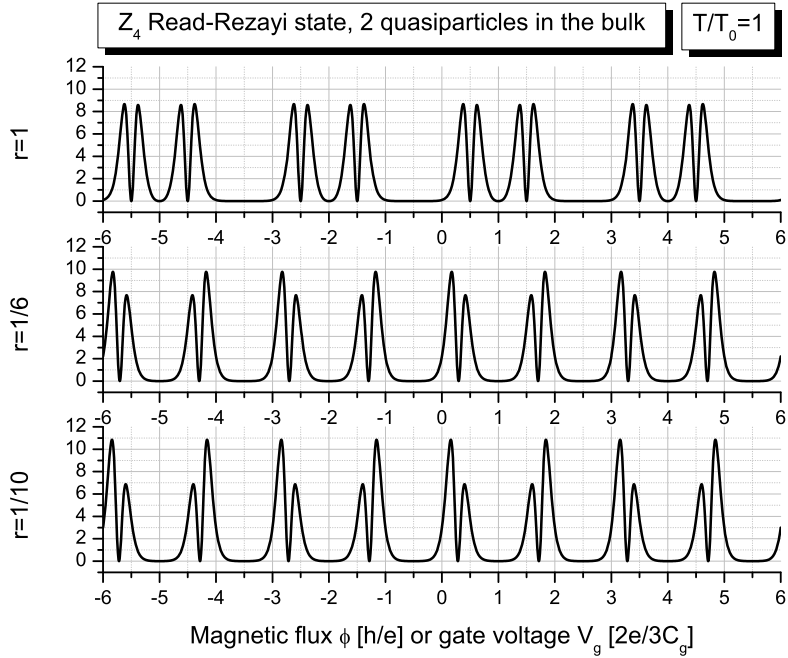}
\caption{Power factors in units [$k_B^2/h$] for a CB island in the $\Z_4$-parafermion (Read-Rezayi) state with $\nu_H=2/3$, 
with $2$ quasiparticles localized  in the bulk, corresponding to $l=2$ and $\rho=1$  in Eq.~(\ref{full-ch}), 
with  $r=1$,  $r=1/6$ and $r=1/10$, at temperature 
$T/T_0=1$. \label{fig:PF-Z4-21-T10-r1-r1-6-r1-10}}
\end{figure}

 Finally we plot in  Fig.~\ref{fig:PF-Z4-21-T10-r1-r1-6-r1-10} the thermoelectric power factor for the $\Z_4$ parafermion
state with two quasiparticles localized in the bulk  corresponding to $l=2$ and $\rho=1$  in Eq.~(\ref{full-ch})
for three different values of $r$.
 Again, the power factor for $r=1$ is symmetric and the sharp dips of $\P_T$ mark the positions of the conductance peaks for all
 values of $r$. There are no peaks of $\P_T$ which remain unchanged when $r$ is decreased and the asymmetries in the power factor
 become more visible when the ratio $r=v_n/v_c$ is decreased.
 
The topological order of the $\Z_4$ parafermion FQH state is $15$ which means that there could be $15$ topologically inequivalent
quasiparticle excitations \cite{NPB2001} in the bulk of the Coulomb blockaded quantum Hall island.
However, the cases with more quasiparticles localized in the bulk are the same as one of the three patterns described in 
Sect.~\ref{sec:Z4_00}, Sect.~\ref{sec:Z4_11}  and Sect.~\ref{sec:Z4_21} up to some translation on the horizontal axis.
This translation may not have physical meaning unless there is a natural way to fix the origin of the horizontal axis.
This completes the description of the thermoelectric power factor in the the $\Z_4$ parafermion Coulomb blockaded islands.
\section{Discussion}
In this paper we gave a complete description of the thermoelectric power factor profiles in the $\Z_3$ and $\Z_4$ 
parafermion states with arbitrary number of quasiparticles localized in the bulk of a Coulomb-blockaded  island. 
The results show that the power factor is rather sensitive to the neutral characteristics of the strongly correlated
two-dimensional electron systems.
While the low-temperature conductance peak patterns in the Coulomb blockade regime for states corresponding to the same 
filling factor $\nu_H$  are practically indistinguishable \cite{nayak-doppel-CB}, for $v_n \ll v_c$ this is so even at finite temperature, 
where $v_n$ and $v_c$ are the Fermi velocities of the neutral and charged modes respectively. 
Surprisingly,  the power factors $\P_T$ of the corresponding states are much more sensitive 
to the neutral modes \cite{NPB2015}.
It appeared that the smaller $r=v_n/v_c$ the bigger the asymmetries in the power factor which combined with 
the thermally induced broadening of the conductance peaks due to the neutral modes' multiplicities could give us the ultimate tool 
to figure out which of the competing quantum Hall universality classes are indeed realized in the experiments. 

 Measuring the thermoelectric power factor, like in Ref.~\cite{gurman-2-3}, that can be computed numerically from the 
thermopower \cite{NPB2015} could  allow us to estimate experimentally the ratio $r=v_n/v_c$ of the Fermi velocities of the 
neutral and charged edge modes.
Notice that this non-universal parameter might depend on the details of the experimental setup and might differ from 
sample to sample, so in any case it is good to have an experimental method to estimate it.
 
The calculated thermoelectric power factors $\P_T$ for the $\Z_k$ parafermion states, as well as those for 
the $\nu_H=5/2$ state \cite{NPB2015},  show that when $r$ is decreased $\P_T$ displays noticeable 
asymmetries in the heights of the $\P_T$ peaks surrounding the conductance peaks positions, which combined with the 
 relatively precise measurement of the ratio $r$, could give some experimental signature for distinguishing different candidates
to describe the experimental data. For some cases of quasiparticles localized in the bulks, some of the peaks of the power factor 
remain unchanged when the ratio $r=v_n/v_c$ is decreased and moreover they are 
characterized by their full symmetry around the  conductance peak position, while the other peaks are strongly asymmetric.

Following the approach of Ref.~\cite{NPB2015}  we propose an experimental way to choose between the different candidates
for the observed $\nu_H=12/5$ Hall state \cite{pan,xia,pan-xia-08}: assume that the profile of the thermoelectric power factor 
for this FQH state has been measured experimentally, e.g., by the methods of Ref.~\cite{gurman-2-3} -- then from the dips of 
the measured power factor we can estimate the value of $r$ and recalculate the theoretical prediction for the power factor 
of the different candidate states for that $r$. After that we can compare the theoretical power factors with the experimental 
one and choose the closest one by the $\chi^2$ method. 

Furthermore, if this method can be combined with controllable deviation from the center of the Hall plateau to simulate 
different number of quasiparticles localized in the bulk that could give us more convincing information about the complete 
structure of the quantum Hall state at filling factor $\nu_H=12/5$. 
\section*{Acknowledgments}
I thank Andrea Cappelli, Guillermo Zemba and Ady Stern for helpful discussions.
This work has been partially supported by the Alexander von Humboldt Foundation under the Return Fellowship and 
Equipment Subsidies Programs and by the Bulgarian Science Fund under Contract No. DFNI-E 01/2 and  DFNI-T 02/6.
\section*{References}
\bibliography{my,TQC,FQHE,Z_k,CB}

\end{document}